\documentclass[aps,prd,reprint,showpacs,amsfonts,amssymb,amsmath]{revtex4-1}

\usepackage[T1]{fontenc}

\usepackage{graphicx}
\usepackage{bm} 
\usepackage[justification=centerlast]{caption}
\usepackage[listofformat=parens]{subfig}

\makeatletter

\usepackage[pdfauthor={Malte F. Linder, Christian Schneider, Joachim Sicking,
Nikodem Szpak, and Ralf Sch\"utzhold},unicode=true, pdfusetitle, bookmarks=true,
bookmarksnumbered=false,bookmarksopen=true,bookmarksopenlevel=1,
breaklinks=false,pdfborder={0 0 1},backref=false,colorlinks=false]{hyperref}

\makeatother

\renewcommand{\Re}{\operatorname{Re}}
\renewcommand{\Im}{\operatorname{Im}}

\DeclareMathOperator{\arsinh}{arsinh}
\DeclareMathOperator{\arcosh}{arcosh}
\DeclareMathOperator{\erf}{erf}
\DeclareMathOperator{\erfi}{erfi}

\newcommand{\ord}{\mathcal{O}}

\newcommand{\keld}{\gamma} 
\newcommand{\action}{\mathcal{A}}

\newcommand{\bea}{\begin{eqnarray}}
\newcommand{\ea}{\end{eqnarray}}
\newcommand{\nn}{\nonumber\\}
\newcommand{\f}[1]{\bm{#1}}

\begin{document}

\title{Pulse shape dependence in the dynamically assisted Sauter-Schwinger
effect}

\author{Malte F. \surname{Linder}}
\author{Christian Schneider}
\author{Joachim Sicking}

\author{Nikodem Szpak}
\email{nikodem.szpak@uni-due.de}

\author{Ralf Sch\"utzhold}
\email{ralf.schuetzhold@uni-due.de}

\affiliation{Fakult\"at~f\"ur~Physik, Universit\"at~Duisburg-Essen,
Lotharstr.~1, 47057~Duisburg, Germany}

\date{October 12, 2015}

\begin{abstract}
While the Sauter-Schwinger effect describes nonperturbative electron-positron
pair creation from vacuum by a strong and slowly varying electric field
$E_{\rm strong}$ via tunneling,
the dynamically assisted Sauter-Schwinger effect corresponds to a strong
(exponential) enhancement of the pair-cre\-a\-tion probability by an additional
weak and fast electric or electromagnetic pulse $E_{\rm weak}$.
Using the WKB and worldline instanton method, we find that this enhancement
mechanism strongly depends on the shape of the fast pulse.
For the Sauter profile $1/\cosh^2(\omega t)$ considered previously,
the threshold frequency $\omega_{\rm crit}$
(where the enhancement mechanism sets in) is basically
independent of the magnitude $E_{\rm weak}$ of the weak pulse---whereas for
a Gaussian pulse $\exp(-\omega^2t^2)$, an oscillating profile $\cos(\omega t)$
or a standing wave $\cos(\omega t)\cos(kx)$,
the value of $\omega_{\rm crit}$ does depend
(logarithmically) on $E_{\rm weak}/E_{\rm strong}$.
\end{abstract}

\pacs{12.20.-m, 
11.15.Tk
}

\maketitle

\section{Introduction}

In contrast to the perturbative realm of quantum field theory, where we 
possess well-established methods for calculating observables such as 
scattering cross sections, our understanding of nonperturbative effects 
is still rather incomplete. 
In quantum electrodynamics (QED), a prominent example for such a 
nonperturbative phenomenon is the Sauter-Schwinger effect
\cite{Sauter,Euler,Weisskopf,Schwinger}
corresponding to the creation of electron-positron pairs out of the 
quantum vacuum by tunneling induced by a strong electric field $E$. 
For constant electric fields $E$, the leading-order pair-cre\-a\-tion 
probability scales as 
\bea
\label{Schwinger-probability}
P_{e^+e^-}
\sim 
\exp\left\{-\pi\,\frac{m^2}{qE}\,\frac{c^3}{\hbar}\right\}
=
\exp\left\{-\pi\,\frac{E_S}{E}\right\}
\,,
\ea
where $\pm q$ is the charge and $m$ the mass of the electrons and 
positrons.
Since the critical field strength $E_S \approx 10^{18}\,\mathrm{V}/\mathrm{m}$
is extremely large, this prediction has not been conclusively 
experimentally verified yet \cite{Experimental-multiphoton-footnote}.

Several years ago, it was found \cite{Assisted} that the above pair-cre\-a\-tion
probability $P_{e^+e^-}$ can be drastically enhanced by superimposing the 
constant (or slowly varying) strong field $E$ with a weaker time-dependent 
pulse---the dynamically assisted Sauter-Schwinger effect; see also Refs.~%
\cite{Ruf-Laser-fields,Hebenstreit-sub-cycle,Catalysis,Monin-Photon-stimulated,
Mocken,Monin+Voloshin-1,Dumlu-Chirped,Dumlu+Dunne-Interference,
Orthaber-Momentum-spectra,Fey,Jiang-Enhancement,Nuriman-Enhanced,
Kohlfuerst-Optimizing,Nuriman-Pulse-shape,Jansen,Dumlu-Resonances,
Blinne-Rotating,Li-Multiple-slit,Strobel+Xue,Akal-Bifrequent,
Hebenstreit-Optimization,Otto,Schneider,Bethe-Heitler}.
Interestingly, this enhancement mechanism is already operative for frequency 
scales $\omega$ far below the mass gap $2mc^2$ separating the Dirac sea from 
the positive continuum. 
In the following, we study the dependence of the dynamically assisted 
Sauter-Schwinger effect on the shape of the additional time-dependent pulse. 
To this end, we employ the WKB and the worldline instanton methods in order 
to compare a Gaussian pulse and an oscillating or standing-wave profile with 
the Sauter pulse considered previously. 

\section{Riccati equation}

Let us briefly review the main steps of the derivation (see e.g.\ Ref.~%
\cite{Dumlu+Dunne-Interference} for a more detailed introduction).
For simplicity, we start with the Dirac equation in 1+1 dimensions 
$(\hbar=c=1)$
\begin{eqnarray}
\label{Dirac}
\left(i\f{\gamma}^\mu[\partial_\mu+iqA_\mu]+m\right)
\cdot\f{\psi}=0
\,,
\end{eqnarray}
with the two-component spinor $\f{\psi}$. 
The purely time-dependent electric field $E(t)$ is treated as an external 
background field and can be described by the vector potential 
$A_\mu=[0,A(t)]$ via $E(t)=\dot A(t)$ in temporal gauge. 
Using the representation of the $\f{\gamma}^\mu$ in terms of the usual 
Pauli matrices $\f{\sigma}_{x,y,z}$, we get the Hamiltonian form
\bea
\label{Dirac-Pauli}
i\partial_t\f{\psi}=\left(-i\f{\sigma}_x\partial_x
+qA(t)\f{\sigma}_x+m\f{\sigma}_z\right)\cdot\f{\psi}
\,.
\ea
After a spatial Fourier transform, this gives
\bea
\label{Dirac-Fourier}
i\partial_t\f{\psi}_k
=
\left([k+qA(t)]\f{\sigma}_x+m\f{\sigma}_z\right)\cdot\f{\psi}_k
=\f{H}_k\cdot\f{\psi}_k
\,,
\ea
where $\f{H}_k$ is a self-adjoint matrix with the eigenvalues
\begin{equation}
\label{eq:Omega-definition}
\f{H}_k\cdot\f{u}_k^\pm
=
\pm\sqrt{m^2+[k+qA(t)]^2}\,\f{u}_k^\pm
=\pm\Omega_k(t)\,\f{u}_k^\pm
\,,
\end{equation}
and real (time-dependent) eigenvectors $\f{u}_k^\pm(t)$ with the
properties $(\f{u}_k^\pm)^2=1$ and $\f{u}_k^+\cdot\f{u}_k^-=0$.

Now we expand the spinor solution $\f{\psi}_k(t)$ of 
Eq.~\eqref{Dirac-Fourier} into these instantaneous eigenvectors
\bea
\label{instantaneous}
\f{\psi}_k(t)
=
\alpha_k(t)e^{-i\varphi_k(t)}\f{u}_k^+(t)
+
\beta_k(t)e^{+i\varphi_k(t)}\f{u}_k^-(t)
\,,
\ea
where we have factored out the rapidly oscillating phase 
\bea
\label{phase-function}
\varphi_k(t)=\int\limits^t_{t_0} dt'\,\Omega_k(t')
\,.
\ea
Inserting the ansatz~\eqref{instantaneous} into the Dirac 
equation~\eqref{Dirac-Fourier} and using 
$\f{\dot u}_k^+\cdot\f{u}_k^+=\f{\dot u}_k^-\cdot\f{u}_k^-=0$, 
we find the evolution equations for the Bogoliubov coefficients 
\bea
\label{Bogoliubov}
\dot\alpha_k(t) &=& +\Xi_k(t)\,\beta_k(t)\, e^{+2i\varphi_k(t)}
\,,
\nn
\dot\beta_k(t) &=& -\Xi_k(t)\,\alpha_k(t)\, e^{-2i\varphi_k(t)}
\,,
\ea
where we have introduced the abbreviation 
\bea
\Xi_k(t)=\f{\dot u}_k^+\cdot\f{u}_k^-=-\f{\dot u}_k^-\cdot\f{u}_k^+=
\frac{mq\dot A(t)}{2\Omega_k^2(t)}
\,.
\ea
Note that $|\alpha_k|^2+|\beta_k|^2$ is conserved and can be set to unity 
$|\alpha_k|^2+|\beta_k|^2=1$ by appropriate initial conditions. 
Introducing the ratio $R_k(t)=\alpha_k(t)/\beta_k(t)$, we can combine the two 
equations~\eqref{Bogoliubov} into one equality 
\bea
\label{Riccati}
\dot R_k(t)
=
\Xi_k(t)\left(e^{2i\varphi_k(t)}+R^2_k(t)e^{-2i\varphi_k(t)}\right)
\,,
\ea
which is the Riccati equation. 
In contrast to the bosonic case (where $|\alpha_k|^2-|\beta_k|^2=1$),
we have a plus sign within the brackets, which reflects the fermionic
nature of the Dirac equation ($|\alpha_k|^2+|\beta_k|^2=1$). 

\section{WKB analysis}

In a stationary situation, the Bogoliubov coefficients $\alpha_k$ and $\beta_k$
in Eq.~\eqref{instantaneous}
correspond to solutions with positive and negative energies, respectively. 
Thus, $\beta_k$ can be interpreted as the amplitude of an electron in the 
Dirac sea while $\alpha_k$ describes an electron in the positive continuum.
Assuming that the field is switched off initially and finally 
$E(t\to\pm\infty)=0$, electron-positron pair creation can be understood 
as the transition of an electron from the Dirac sea into the positive 
continuum induced by the electric field $E(t)$.
Thus, with the initial conditions $\alpha_k^{\rm in}=0$ and
$\beta_k^{\rm in}=1$, the pair-cre\-a\-tion probability (for the mode $k$) is
given by 
\bea
P^{e^+e^-}_k=\left|\alpha_k^{\rm out}\right|^2=
\frac{\left|R_k^{\rm out}\right|^2}{\left|R_k^{\rm out}\right|^2+1}
\,.
\ea
Note that this picture is a bit different from the usual description of 
particle creation in the bosonic case, where the probability is given by 
$|\beta_k^{\rm out}|^2$ instead. 
However, for fermions, the particle-hole duality 
(which is the basis for the Dirac sea construction) implies that 
$\alpha_k$ and $\beta_k$ behave symmetrically.
Thus, one can use both representations (Dirac sea filled with either 
electrons or positrons) with the correct initial conditions. 

Since this probability $P^{e^+e^-}_k$ is realistically very small, 
we may approximate $|\alpha_k^{\rm out}|\approx|R_k^{\rm out}|\ll1$. 
For solutions with small amplitude $R_k(t)\ll1$, the Riccati 
equation~\eqref{Riccati} can be approximated by
\begin{equation}
\label{Riccati-approx}
\dot R_k(t)\approx\Xi_k(t)e^{2i\varphi_k(t)}
\,.
\end{equation}
Note that this approximation does in general not yield the exact prefactor
\cite{Davis}, but it does give the correct exponent---which is the quantity we
are interested in.

Equation~\eqref{Riccati-approx} can now be integrated with the initial
condition $R_k^{\rm in}=R_k(t\to-\infty)=0$
\begin{equation}
\label{eq:contour-integral}
R_k^{\rm out}=R_k(t\to \infty)
\approx\int\limits_{-\infty}^{+\infty}dt\,\Xi_k(t)\, e^{2i\varphi_k(t)}
\,.
\end{equation}
Assuming that $A(t)$ and thus $\Xi_k(t)$ as well as $\varphi_k(t)$ are 
analytic functions in a strip of the complex plane including the real axis, 
the above integral can be calculated by deforming the 
integration contour into the complex $t$ plane.
In the upper complex half-plane $\Im(t)>0$, the exponential $e^{2i\varphi_k(t)}$
decays rapidly---which corresponds to the fact that $R_k^{\rm out}$ 
is exponentially suppressed. 
However, we cannot deform the integration contour arbitrarily far since 
there will be singularities, e.g., at 
\bea
\label{singularity}
\Omega_k(t_k^\star)=0
\,\leadsto\,
k+qA(t_k^\star)=\pm im 
\,.
\ea
The solution of Eq.~\eqref{Riccati-approx} can thus be estimated by 
\begin{equation}
\label{Ricatti-estimate}
R_k^{\rm out}\sim\exp\left\{2i\varphi_k(t_k^\star)\right\}
\sim\exp\left\{-2\Im[\varphi_k(t_k^\star)]\right\}
\,.
\end{equation}
Let us consider the most simple example: a constant electric field 
$E=\rm const$. 
In this case, we have $A(t)=Et$ and thus we only get two singularities 
\bea
\label{constant}
t_k^\star=\frac{\pm im-k}{qE}
\,.
\ea
Insertion into the estimate~\eqref{Ricatti-estimate} then yields the usual 
Schwinger exponent in Eq.~\eqref{Schwinger-probability}.

In the more general case with more singularities $t_k^\star$, 
they all contribute to the estimate~\eqref{Ricatti-estimate} but typically 
the one with the smallest $|\Im[\varphi_k(t_k^\star)]|$ dominates
(for a more detailed discussion, cf.\ Appendix~\ref{app:contour-integral}).
It can happen, however, that two (or more) have comparable magnitude---in
which case there can be interference effects, which have already been 
observed in Refs.~\cite{Hebenstreit-sub-cycle,Dumlu+Dunne-Interference,
Orthaber-Momentum-spectra,Fey,Dumlu+Dunne-complex,Akkermans+Dunne-Ramsey},
for example.

\section{Sauter profile}

Let us first apply the method sketched above to the case of a Sauter pulse 
\cite{Sauter} considered previously in Refs.~\cite{Assisted,Fey}.
More specifically, we consider an electric field of the form 
\bea
E(t)=E_1+\frac{E_2}{\cosh^2(\omega_2t)}
\,,
\ea
where $E_2\ll E_1\ll E_S$ and $\omega_2\ll m$. 
The associated vector potential reads 
\bea
\label{vector-potential-tanh}
A(t)=E_1t+\frac{E_2}{\omega_2}\tanh(\omega_2t)
\,,
\ea
and has poles at $\omega_2t=i\pi/2+i\pi\mathbb Z$. 
In terms of the dimensionless quantities $\tau=\omega_2t$, 
$\varepsilon=E_2/E_1$, $p=k/m$, as well as the combined Keldysh parameter 
\cite{Assisted,Keldysh}
\bea
\label{Keldysh}
\keld=\frac{m\omega_2}{qE_1}
\,,
\ea
the equation~\eqref{singularity} for the singularities reads 
\bea
\tau^\star+\varepsilon\tanh\tau^\star=\keld(\pm i-p) 
\,.
\ea
Consequently, for small $\varepsilon$, we basically get the same 
singularity~\eqref{constant} in the upper half-plane as for a constant field, 
which we call the regular singularity 
\bea
\label{regular-singularity}
\tau^\star_{\rm reg}=\keld(i-p)+\ord(\varepsilon) 
\,.
\ea
However, from the analytic continuation of the vector 
potential~\eqref{vector-potential-tanh}, we get additional singularities 
from the poles of the complex $\tanh$ function, for example at  
\bea
\label{additional-singularity}
\tau^\star_{\rm add}=\frac{\pi}{2}\,i 
\,.
\ea
Now, setting $p=0$ for simplicity, we have the following picture:
For small Keldysh parameters {[}Eq.~\eqref{Keldysh}{]}, the regular singularity
is closer to the real axis and thus its contribution~\eqref{Ricatti-estimate}
dominates. 
In this case, we obtain basically the same exponent as for the constant 
strong field $E_1$ alone (ordinary Sauter-Schwinger effect). 
If the Keldysh parameter exceeds the critical (threshold) value of 
\bea
\label{critical-Keldysh}
\keld_{\rm crit}=\frac{\pi}{2}
\,,
\ea
however, the additional singularity $\tau^\star_{\rm add}=i\pi/2$ caused by the
Sauter pulse $E_2/\cosh^2(\omega_2t)$ is closer to the real axis and thus 
its contribution starts to dominate.
Thus, the exponent in Eq.~\eqref{Ricatti-estimate} is reduced and hence the 
probability enhanced (dynamically assisted Sauter-Schwinger effect). 
Inserting the singularity $\tau^\star$ into the 
estimate~\eqref{Ricatti-estimate} yields 
\bea
\Im\left[\varphi_k(t_k^\star)\right]\approx\frac{E_S}{2E_1}\,
\Im\left[\phi\!\left(\frac{\tau^\star}{\keld}+p\right)\right]
\,,
\ea
with the auxiliary function \cite{Assisted,Fey}
\bea
\label{eq:Schwinger-phase-aux-func}
\phi(z)=z\sqrt{1+z^2}+\arsinh z
\,.
\ea
For $\keld<\keld_{\rm crit}$, the regular singularity 
$\tau^\star_{\rm reg}\approx\keld(i-p)$ dominates and we get $\phi(i)=i\pi/2$
which reproduces Eq.~\eqref{Schwinger-probability}.
For $\keld>\keld_{\rm crit}$, on the other hand, we have to insert 
$\tau^\star_{\rm add}=i\pi/2$ which yields the results in Ref.~\cite{Assisted}.
For Keldysh parameters near the critical value~\eqref{critical-Keldysh} 
the competition between regular and additional singularities leads to 
$p$-dependent interference effects \cite{Orthaber-Momentum-spectra,Fey}.

\section{Gauss pulse}

Now let us repeat the same analysis for another bell-shaped curve which is 
visually almost indistinguishable from the Sauter profile 
$1/\cosh^2(\omega_2t)$: a Gauss pulse 
\bea
E(t)=E_1+E_2\exp\{-(\omega_2t)^2\}
\,,
\ea
with the same hierarchy as before, i.e., 
$E_2\ll E_1\ll E_S$ and $\omega_2\ll m$
(see Ref.~\cite{Nuriman-Enhanced} for some numeric results).
The vector potential can be expressed in terms of the error function 
%
\bea
A(t)=E_1t+\frac{\sqrt{\pi} E_2}{2\omega_2}\erf(\omega_2t)
\,.
\ea
As an important difference to the Sauter profile, the Gauss pulse and the 
error function are analytic in the entire complex plane and thus do not give 
rise to additional singularities. 
Hence the only singularities are given by Eq.~\eqref{singularity} 
\bea
\label{eq:Gauss-pulse-sing-eq}
\tau^\star+\varepsilon\,\frac{\sqrt{\pi}}{2}\erf\tau^\star
=\keld(\pm i-p) 
\,.
\ea
Again, for small and moderate values of $\keld$, we basically obtain the 
same regular singularity as in Eqs.~\eqref{constant} and 
\eqref{regular-singularity}
and thus the same exponent as for the constant strong field $E_1$ alone 
(ordinary Sauter-Schwinger effect). 
Moreover, there appears an infinite set of additional singularities
further away from the real axis whose contributions are therefore suppressed.

If $\keld$ becomes large enough, however, the regular singularity starts to 
deviate from its ordinary position \eqref{regular-singularity} while its 
imaginary part approaches that of the additional singularities
(for more details, cf.\ Appendix~\ref{sub:Gauss-pulse-singularities}).
Setting $p=0$ for simplicity, the asymptotic behavior of the error 
function for large imaginary arguments~\cite{Abramowitz+Stegun},
\bea
\erf(i\keld)\sim\frac{\exp\{\keld^2\}}{\keld\sqrt{\pi}}
\,,
\ea
implies that this motion begins when $\exp\{\keld^2\}$ is large enough to 
compensate the smallness of $\varepsilon$, i.e., the critical value of the 
Keldysh parameter roughly scales as 
\bea
\label{critical-Keldysh-Gauss}
\keld_{\rm crit}\sim\sqrt{|\ln\varepsilon|}
\,.
\ea
As a result, although the Gauss pulse looks very similar to the Sauter 
pulse, they behave remarkably differently regarding the dynamically assisted 
Sauter-Schwinger effect.
In one case (Sauter pulse), the critical frequency where the enhancement 
sets in is basically independent of $\varepsilon=E_2/E_1$ whereas in the 
other case (Gaussian), this threshold frequency scales as 
$\omega_2^{\rm crit}\sim\sqrt{|\ln\varepsilon|}$. 

Apart from the motion of the singularities, the additional Gaussian pulse 
does also affect the phase function $\varphi_k$ in 
Eqs.~\eqref{phase-function} and \eqref{Ricatti-estimate}. 
Again setting $p=0$ for simplicity, we get to first order in $\varepsilon$
\begin{equation}
\label{WKB-Gauss}
\Im\left[\varphi_k(t_k^\star)\right]\approx\frac{\pi E_S}{4E_1}
\left[1-\varepsilon e^{\keld^2/2}
\left(
I_0\!\left[\frac{\keld^2}{2}\right]-
I_1\!\left[\frac{\keld^2}{2}\right]
\right)
\right] 
\,,
\end{equation}
where $I_\nu$ are the modified Bessel functions of the first kind. 
Note, however, that this expansion is only reliable for Keldysh parameters 
well below the critical value~\eqref{critical-Keldysh-Gauss}.  

\section{Oscillation}

As our third example, let us consider a harmonically oscillating weak field 
(see also Refs.~\cite{Otto,Akal-Bifrequent,Jansen,Mocken,Jiang-Enhancement,
Nuriman-Enhanced})
\bea
\label{eq:oscillation-E-field}
E(t)=E_1+E_2\cos(\omega_2t) 
\,,
\ea
again with $E_2\ll E_1\ll E_S$ and $\omega_2\ll m$. 
Similar to the Gaussian case, the vector potential 
\bea
\label{eq:oscillation-potential}
A(t)=E_1t+\frac{E_2}{\omega_2}\,\sin(\omega_2t) 
\,,
\ea
is analytic in the entire complex plane.
Again Eq.~\eqref{singularity} yields the singularities 
\bea
\label{eq:oscillation-sing-eq}
\tau^\star+\varepsilon\,\sin\tau^\star
=\keld(\pm i-p) 
\,.
\ea
The solutions of this transcendental equation can be determined graphically
(cf.\ Appendix~\ref{sub:App-singularities-osc}).
If we again set $p=0$ for simplicity, the dominant, regular singularity 
is purely imaginary and given by 
\bea
\label{eq:osc-zerop-main-sing-eq}
\Im\tau^\star+\varepsilon\sinh(\Im\tau^\star)=\keld
\,.
\ea
As in the Gaussian case, small and moderate values of $\keld$ basically do
not affect the singularity (since $\varepsilon\ll1$) and the dynamically 
assisted Sauter-Schwinger effect requires $\keld$ to be large enough 
such that the magnitude of $\sinh\keld$ compensates the smallness of 
$\varepsilon$.
In this limit, we may approximate $\sinh\keld\approx e^\keld/2$ and thus 
the critical value of the Keldysh parameter roughly scales as 
\bea
\label{critical-Keldysh-cos}
\keld_{\rm crit}\sim|\ln\varepsilon|
\,.
\ea
Again, we may Taylor expand the phase function $\varphi_k$ in 
Eqs.~\eqref{phase-function} and \eqref{Ricatti-estimate} 
for Keldysh parameters well below this critical 
value~\eqref{critical-Keldysh-cos} 
\bea
\label{WKB-cos}
\Im\left[\varphi_k(t_k^\star)\right]\approx\frac{\pi E_S}{4E_1}
\left[1-2\varepsilon\cos(p\keld)\,\frac{I_1(\keld)}{\keld}  
\right] 
\,. 
\ea
%

\section{Worldline instanton method}

It is illuminating to address the above scenarios with another technique,
the worldline instanton method \cite{Dumlu+Dunne-complex,Feynman,Afflek,
Kim+Page-02,Dunne+Schubert,Dunne,Kim+Page-06,Dunne-overview-08,
Schubert-Lectures}.
In contrast to the WKB approach, the worldline instanton technique does not 
yield the momentum dependence (it only provides the total pair-cre\-a\-tion 
probability), but it has the advantage that some calculations are easier to do.
Again, let us begin with a brief review of this method.
The total pair-cre\-a\-tion probability $P_{e^+e^-}$ is related to the 
vacuum persistence amplitude $\left<0_\mathrm{out}|0_\mathrm{in} \right>$ 
\begin{equation}
P_{e^+e^-} = 1 - \left|\left<0_\mathrm{out}|0_\mathrm{in} \right>\right|^2
\,.
\end{equation}
If we now express the vacuum persistence amplitude 
$\left<0_\mathrm{out}|0_\mathrm{in} \right>$ in terms of the path integral 
and perform a series of transformations and approximations, we find that 
$P_{e^+e^-}$ can be estimated in analogy to Eq.~\eqref{Ricatti-estimate} via 
\begin{equation}
\label{instanton-estimate}
P_{e^+e^-} \sim e^{-\action}
\,.
\end{equation}
Here $\action$ is the worldline instanton action in the presence of the 
electric field $E$ represented by the Euclidean vector potential 
$A_\mu(x^\nu)$ after analytic continuation to imaginary time 
$x_\mu=(x,y,z,it)=(x_1,x_2,x_3,x_4)$
\begin{equation}
\label{eq:instantonaction}
\action = ma + iq \int\limits_0^1 du\,\dot x^\mu A_\mu(x^\nu[u])
\,.
\end{equation}
The worldline instanton $x_\mu(u)$ is a closed loop in Euclidean 
space-time satisfying the boundary conditions $x_\mu(0)=x_\mu(1)$ 
and the instanton equations of motion
\begin{equation}
\label{eq:instantoneqs}
m\ddot{x}_\mu = i q a F_{\mu\nu} \dot{x}_\nu
\,,
\end{equation}
with $\dot{x}_\mu=dx_\mu/du$ and $\ddot{x}_\mu=d^2x_\mu/du^2$.
Since $F_{\mu\nu}$ is antisymmetric, the absolute value of the 
``four-velocity'' 
\bea
\label{four-velocity}
\dot{x}_\mu\dot{x}^\mu
=
\dot{x}_1^2 + \dot{x}_2^2+\dot{x}_3^2 + \dot{x}_4^2
=
a^2
\,,
\ea
is conserved. 
%

\subsection{\label{sec:intform}Integral form of the instanton action}

\begin{figure}
\includegraphics[width=0.8\linewidth]{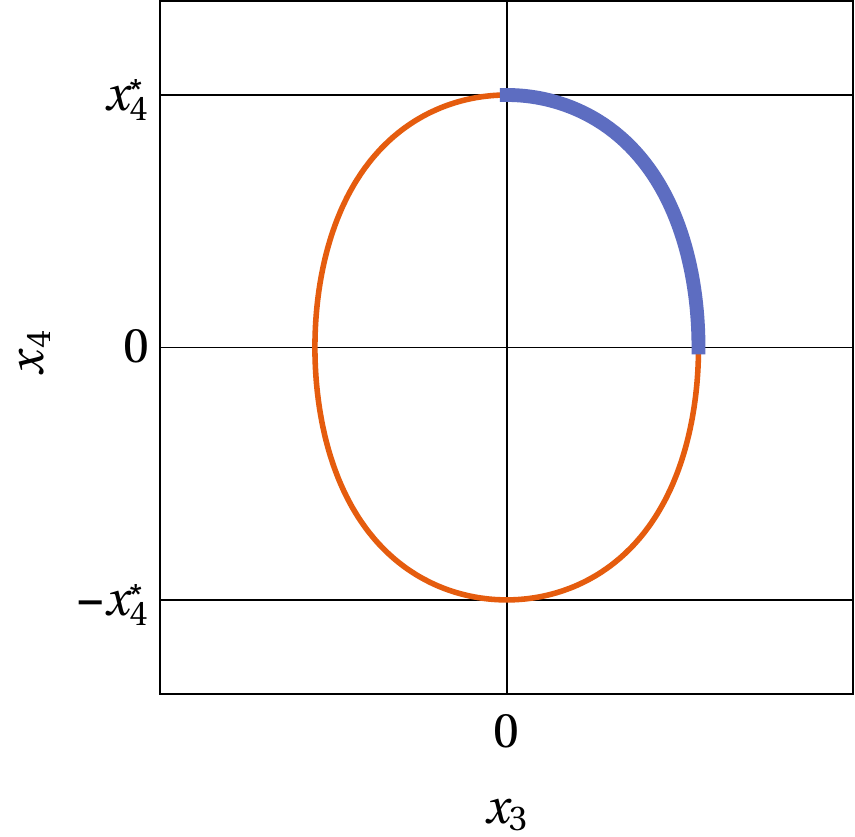}%
\caption{\label{fig:trajectory}Example of an instanton trajectory 
$x_\mu(u)$ with the appropriate symmetries. 
In this case, the full instanton action $\action$ is four times the 
action along the highlighted (or any) quarter of the trajectory.}
\end{figure}

In certain cases, for example a purely time-dependent electric field 
$\f{E}=E(t)\f{e}_z$, the instanton action $\action$ can be computed in an
especially simple way. 
Let us consider a Euclidean four-potential of the form
\begin{equation}
\label{eq:potential}
i A_3(x_4)=\frac{E}{\omega}\,f(x_4\omega)
\,,
\end{equation}
where $f(z)$ is a dimensionless odd analytic function such that the electric 
field $\f{E}=f'(i\omega t)\f{e}_z$ is real. 
Inserting the vector potential \eqref{eq:potential}, the only nonvanishing 
components of $F_{\mu\nu}$ are $iF_{43} = -iF_{34} = Ef'(\omega x_4)$ and the 
instanton equations~\eqref{eq:instantoneqs} reduce to
\begin{subequations}
\begin{alignat}{2}
\label{eq:insteq3}\ddot{x}_3 &= -&\frac{q E a}{m} f'(\omega x_4)\dot{x}_4
\,, 
\\
\label{eq:insteq4}\ddot{x}_4 &=  +&\frac{q E a}{m} f'(\omega x_4)\dot{x}_3
\,,
\end{alignat}
as well as $\ddot{x}_1=\ddot{x}_2=0$.
Thus, in order to have closed loops $x_\mu(0)=x_\mu(1)$, we must have 
$\dot{x}_1=\dot{x}_2=0$ and hence Eq.~\eqref{four-velocity} simplifies to 
\begin{equation}
\label{eq:insteqa}
a^2 = \dot{x}_3^2 + \dot{x}_4^2
\,.
\end{equation}
\end{subequations}
Equation~\eqref{eq:insteq3} can be integrated immediately to give
\begin{equation}
    \label{eq:dx3}
    \dot{x}_3 = - \frac{a}{\keld}\, f(\omega x_4)\,,
\end{equation}
where we have set the integration constant to zero in order 
to get closed loops. 
Using \eqref{eq:insteqa}, we arrive at
\begin{equation}
\label{eq:dx4}
\dot{x}_4
=
\pm\sqrt{a^2-\dot{x}_3^2}
=
\pm a\sqrt{1-\frac{1}{\keld^2}f^2(\omega x_4)}
\,.
\end{equation}
Furthermore, for a field of the form~\eqref{eq:potential}, the instanton 
action~\eqref{eq:instantonaction} can be simplified as well
\begin{align}
\action 
&\overset{\eqref{eq:potential}}{=}
ma + \frac{qE}{\omega} \int\limits_0^1 du\, f(\omega x_4) \dot{x}_3 
\nonumber\\
&\overset{\eqref{eq:dx3}}{=}
ma - \frac{m}{a} \int\limits_0^1 du\, \dot{x}_3^2
\overset{\eqref{eq:insteqa}}{=}
\label{eq:actionx4}
\frac{m}{a} \int\limits_0^1 du\,\dot{x}_4^2
\,.
\end{align}
Since $f(x_4\omega)$ is an odd function and thus $f'(x_4\omega)$ and 
$f^2(x_4\omega)$ are even functions, the instanton trajectory has a fourfold 
symmetry: $x_3\to-x_3$ and $x_4\to-x_4$. 
As a result, the contribution to the action~\eqref{eq:actionx4} is the same 
for each quarter of the trajectory; see Fig.~\ref{fig:trajectory}. 
In each of these quarters, $\dot{x}_4$ is a unique function
{[}Eq.~\eqref{eq:dx4}{]} of $x_4$ and thus we can turn the
$u$ integral~\eqref{eq:actionx4} into an $x_4$ integral
\begin{align}
\action 
&=
4 \frac{m}{a}\int\limits_0^{1/4}du\, \dot{x}^2_4 
\overset{\eqref{eq:dx4}}{=}
4 m \int\limits_{x_4(0)}^{x_4(1/4)} dx_4\,
\sqrt{1-\frac{1}{\keld^2}f^2(\omega x_4)} 
\nonumber\\
\label{eq:intaction}
&= 
4 \frac{m^2}{qE} \int\limits_0^{\chi^\star} d\chi\,
\sqrt{1-\frac{1}{\keld^2}f(\keld\chi)^2}\,,
\end{align}
where we have substituted $\chi=qEx_4/m$ in the last line.
The integral coincides (up to substitutions) with the function $g(\gamma^2)$
in Ref.~\cite{Dunne}.
The turning point $\chi^\star$ is given implicitly by the condition
\begin{equation}
\keld = f(\keld\chi^\star)
\,,
\end{equation}
which corresponds to $t_k^\star$ in Eq.~\eqref{singularity} for $p=0$,
where we have $\tau^\star=i\keld\chi^\star$. 
The expression~\eqref{eq:intaction} is generally much easier to evaluate 
numerically---and approximate analytically---than solving the equations of 
motion~\eqref{eq:instantoneqs}.

\section{Dynamical assistance}

\begin{figure}
\includegraphics[width=0.85\linewidth]{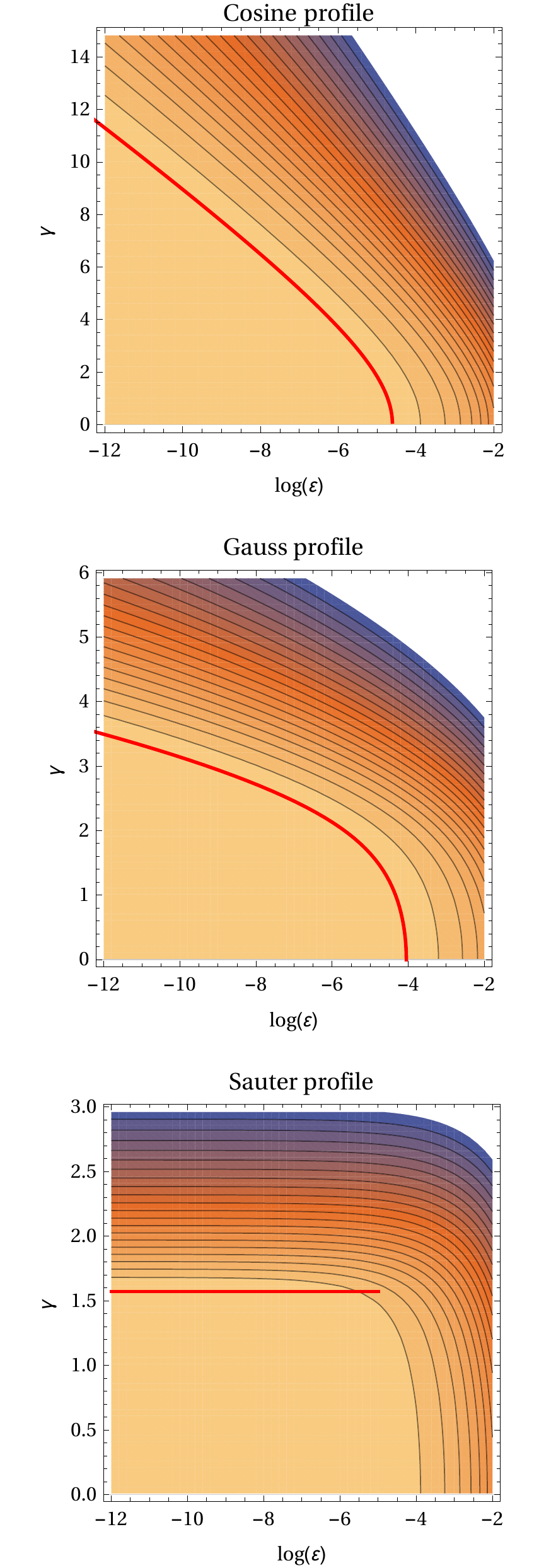}%
\caption{\label{fig:profiles}Instanton action $\action$ ranging from
$\pi E_S/E$ (light yellow) to $2 E_S/E$ (dark blue) for the considered 
profiles as a function of $\keld$ and $\varepsilon$. 
This thin (black) lines are contour lines and 
the additional thick (red) lines depict the corresponding 
estimates~\protect\eqref{eq:cosineScaling}, \protect\eqref{gauss-linear},
and \protect\eqref{critical-Keldysh}, respectively.
Note the different scales of the $\keld$ axes.}
\end{figure}

In order to study the dynamically assisted Sauter-Schwinger effect described
above, we choose the dimensionless function $f(\chi)$ from
Sec.~\ref{sec:intform} to be
\begin{equation}
f(\chi) = \chi + \varepsilon g(\chi) 
\,,
\end{equation}
with $\varepsilon \ll 1$ and a function $g(\chi)$ representing the 
additional weak time-dependent profile
\begin{equation}
\f{E}=E\left[1 + \varepsilon g'(i \omega t)\right]\f{e}_z
\,.
\end{equation}
Assuming that $g(\chi)$ is a well-behaved function 
(which is the case for the Gaussian and the oscillating profile---but not for
the Sauter pulse due to its singularities), we can expand the
integral~\eqref{eq:intaction} to first order in $\varepsilon$
\begin{align}
\action 
&= 
4 \frac{m^2}{qE} \int\limits_0^{x^*} d\chi\; 
\sqrt{1-\left(\chi + \varepsilon\,\frac{g(\keld\chi)}{\keld}\right)^2}
\nonumber\\
&\approx 
4\frac{m^2}{qE} \left(\int\limits_0^1 d\chi \sqrt{1-\chi^2}
- \varepsilon \int\limits_0^1\frac{d\chi\;\chi}{\sqrt{1-\chi^2}}\, 
\frac{g(\keld\chi)}{\keld}\right)
\nonumber\\
\label{eq:actionReduced}
&= 
\frac{m^2\pi}{qE}
\left(1 - \frac{4 \varepsilon}{\pi \keld}
\int\limits_0^1 d\xi\, g\left(\keld \sqrt{1-\xi^2}\right)
\right)
\nonumber\\
&= 
\pi\,\frac{E_S}{E}\left(1 - \frac{4 \varepsilon}{\pi\keld}\,G(\keld)\right) 
\,.
\end{align}
To first order in $\varepsilon$, the dependence of $\chi^\star$ on 
$\varepsilon$ does not contribute because the integrand vanishes at that 
point. 
The contribution to zeroth order in $\varepsilon$ reproduces the 
constant field in Eq.~\eqref{Schwinger-probability}.
The form of $G(\keld)$ then determines how much, for an additional pulse
$\varepsilon g'(i \omega t)$, a given strength $\varepsilon$ and Keldysh 
parameter $\keld$, the instanton action $\action$ is reduced and thus pair 
production improved.

\subsection{Gaussian pulse}

For a Gaussian pulse, the Euclidean vector potential is represented by the 
imaginary error function 
\bea
g(\chi)=\frac{\sqrt{\pi}}{2}\,\erfi\chi
\,,
\ea
and thus the function $G(\keld)$ in Eq.~\eqref{eq:actionReduced} reads 
\bea
\label{gauss-linear}
G(\keld)
=
\frac{\pi\keld}{4}\,e^{\keld^2/2}
\left(
I_0\left[\frac{\keld^2}{2}\right]-
I_1\left[\frac{\keld^2}{2}\right]
\right)
\,.
\ea
Comparison with Eq.~\eqref{WKB-Gauss} shows agreement for $p=0$, which 
demonstrates that the pairs are produced predominantly with $p=0$
(as expected). 
Of course, this is no accidental coincidence as the manipulations in 
Eq.~\eqref{eq:actionReduced} are equivalent to those leading to 
Eqs.~\eqref{WKB-Gauss} and \eqref{WKB-cos} for $p=0$; see also 
Ref.~\cite{Strobel+Xue}.

For large $\keld$, the function $G(\chi)$ behaves as 
$\exp\{\keld^2\}/\keld^2$ and thus the threshold for the 
dynamically assisted Sauter-Schwinger effect is reached when  
\bea
\varepsilon\sim\keld^3 e^{-\keld^2}
\,,
\ea
which is consistent with Eq.~\eqref{critical-Keldysh-Gauss}. 
Again, we would like to stress that the first-order 
expansion~\eqref{gauss-linear} is only valid for Keldysh parameters 
$\keld$ well below this critical value. 

\subsection{Oscillating profile}

In order to represent an oscillation, we choose 
\bea
g(\chi)=\sinh\chi
\,.
\ea
In this case, the $G(\keld)$ in Eq.~\eqref{eq:actionReduced} becomes 
\bea
\label{eq:cosineScaling}
G(\keld)=\frac{\pi}{2}\,I_1(\keld)
\,.
\ea
Again, comparison with Eq.~\eqref{WKB-cos} shows agreement for $p=0$ 
as well as for $p=\pm2\pi/\keld$ and $p=\pm4\pi/\keld$ etc. 
These additional $p$ values result from the periodicity of the oscillation 
which implies that the times $t$ and $t\pm T$ with $T=2\pi/\omega$ are 
equivalent.
During one period $T$, the acceleration by the strong electric field $E$ 
causes a momentum shift of $\Delta k=qET=qE2\pi/\omega=2\pi m/\keld$,
i.e., $\Delta p=2\pi/\keld$.  

In view of the asymptotic behavior $G(\keld)\sim\keld^{3/2} e^{-\keld}$ 
for large $\keld$, the threshold for the dynamically assisted 
Sauter-Schwinger effect is reached when
\bea
\varepsilon\sim\keld^{3/2} e^{-\keld}
\,, 
\ea
which is again consistent with Eq.~\eqref{critical-Keldysh-cos}. 

\section{Standing wave}

One might perhaps object that the fields $\f{E}(t)$ considered here are not 
solutions of the Maxwell equations in vacuum, i.e., without sources. 
This objection could be motivated by the idea that the fermionic modes 
somehow ``feel'' the sources generating the electromagnetic background 
field and thus behave totally different in field configurations with and 
without sources. 
However, retracing the steps of the formalism presented above, we see that 
we only used the Dirac equation in a given external background field $A_\mu$.
Whether this field $A_\mu$ is a solution of the Maxwell or Proca or 
Yang-Mills equations with or without sources is irrelevant for our 
derivation. 
Note that we also did not exploit local gauge invariance.  

In order to illuminate this point, let us consider the following 
electric field profile:
\begin{equation}
\label{standing-wave}
\f{E}(t,x) = \left[E_1 + E_2 \cos(\omega t)\cos(k x)\right]\f{e}_z
\,.
\end{equation}
It can be represented by the Euclidean vector potential
\begin{equation}
i A_3 (x_1,x_4) = E_1 x_4 + \frac{E_2}{\omega}\sinh(\omega x_4)\cos(k x_1)
\,,
\end{equation}
which generates the above electric field in $z$ direction as well as 
a magnetic field in the $y$ direction.
Thus, the only nonvanishing components of $F_{\mu\nu}$ are
\begin{subequations}
\begin{align}
i F_{43} &= -iF_{34} = E_1 + E_2 \cosh(\omega x_4) \cos(k x_1)
\,, 
\\
i F_{13} &= -iF_{31} = - E_2\, \frac{k}{\omega}\, \sinh(\omega x_4) \sin(k x_1)
\,,
\end{align}
\end{subequations}
which yields the following instanton equations:
\begin{subequations}
\begin{align}
m\ddot{x}_1 =& -qaE_2\frac{k}{\omega}\sinh(\omega x_4)\sin(k x_1)\dot{x}_3 
\,,
\label{eq:xtrans} 
\\
m\ddot{x}_3 =& -qa\left[E_1 + E_2 \cosh(\omega x_4)\cos(k x_3)\right]\dot{x}_4
\nonumber\\
& + qaE_2\frac{k}{\omega}\sinh(\omega x_4)\sin(kx)\dot{x}_1
\,, 
\\
m\ddot{x}_4 =& qa\left[E_1 + E_2 \cosh(\omega x_4)\cos(k x_3)\right] \dot{x}_3
\,.
\end{align}
\end{subequations}
The equation~\eqref{eq:xtrans} for the transverse component $x_1$ is solved 
by $x_1(u)=0$ or $x_1(u)=\pm\pi/k$ etc. 
Choosing a solution in the maximum of the electric field such as $x_1(u)=0$, 
the other equations reduce to
\begin{subequations}
\begin{align}
m\ddot{x}_3 &= -qa\left[E_1 + E_2 \cosh(\omega x_4)\right]\dot{x}_4
\,, 
\\
m\ddot{x}_4 &= qa\left[E_1 + E_2 \cosh(\omega x_4)\right]\dot{x}_3
\,.
\end{align}
\end{subequations}
As a result, we find that the instanton trajectories and thus also the 
instanton action $\action$ are not affected by the additional factor 
$\cos(k x)$ in Eq.~\eqref{standing-wave} at all.
This invariance can be generalized to vector potentials $A_3(x_1,x_4)$ 
for which $\partial A_3/\partial x_1$ vanishes at $x_1=0$ for all $x_4$, 
which is the case for all even functions $A_3(-x_1,x_4)=A_3(x_1,x_4)$,
for example.  

Previous experience suggests that temporal variations of the electric field 
$E(t)$ tend to increase the pair-cre\-a\-tion probability whereas longitudinal 
spatial variations $E(x)\f{e}_x$ tend to decrease it
(an interplay between these two tendencies is discussed in
Ref.~\cite{Schneider}; see also Ref.~\cite{Ilderton}).
Our example shows that a transversal spatial dependence does not necessarily 
have this effect because the worldline instanton action $\action$ is actually 
independent of $k$. 
Note, however, that the situation would change drastically for a 
propagating wave such as $\cos(\omega t-kx)$ instead of a standing wave 
$\cos(\omega t)\cos(k x)$; see, e.g., Ref.~\cite{Catalysis}.
Furthermore, the prefactor in front of the exponential in 
Eq.~\eqref{instanton-estimate}, which can be derived via studying 
small perturbations around the instanton trajectory, could well depend 
on $k$ since it affects the effective pair-cre\-a\-tion volume.

For $\omega=k$, the field~\eqref{standing-wave} represents a vacuum solution
of the Maxwell equations in the form of a standing wave, which can be generated
by the superposition of two plane waves $\cos(\omega t\pm kx)$ with the same 
polarization but propagating into opposite directions.
For $\omega\neq k$, however, this field profile only solves the Maxwell 
equations with nonzero sources.
Nevertheless, the instanton action $\action$ is the same for $\omega=k$ and 
$\omega\neq k$, which demonstrates that it is irrelevant whether the background
field $A_\mu$ is a vacuum solution of the Maxwell equations or not.
The difference between a standing $\cos(\omega t)\cos(k x)$ and a propagating 
wave $\cos(\omega t-kx)$ is far more important than the question of whether 
the field is a vacuum solution of the Maxwell equations ($\omega=k$) 
or not ($\omega\neq k$).

Of course, we do not deny that it is certainly desirable to calculate the 
pair-cre\-a\-tion probability for field configurations which are as realistic 
as possible.
On the other hand, our results show that one should not discard results just 
because the considered field profile is not a vacuum solution of the Maxwell 
equations \cite{Zahn-footnote}.

\section{Conclusions}

Via the WKB and the worldline instanton method, we studied and compared
the dynamically assisted Sauter-Schwinger effect for three profiles
(Sauter, Gauss, and oscillating) and found qualitative differences.
For the Sauter pulse considered previously, the critical (threshold) Keldysh
parameter~\eqref{critical-Keldysh} is determined by the competition between
the regular singularity~\eqref{regular-singularity} and the additional
singularity~\eqref{additional-singularity} and is basically independent of
the magnitude $E_2$ of the weak field.
By contrast, the other two cases do not feature such a competition between 
different singularities and the critical (threshold) Keldysh 
parameters~\eqref{critical-Keldysh-Gauss} and \eqref{critical-Keldysh-cos}
do depend on the magnitude of the weak field, albeit only logarithmically.
These profound differences can be traced back to the distinct features of the 
three profiles in the complex $t$ plane---which also determines the momentum 
dependence including interference effects etc.

Furthermore, one would expect that these differences do also affect the 
behavior of the prefactor (which was not considered here) in front of the 
exponential in Eqs.~(\ref{Ricatti-estimate}) and \eqref{instanton-estimate},
e.g., its dependence on $\varepsilon$.
Note, however, that this prefactor cannot resolve or counterbalance the 
observed difference between the pulse profiles (e.g., Sauter and Gauss).
For example, let us consider a Keldysh parameter a bit above the threshold
value \eqref{critical-Keldysh} of $\keld_{\rm crit}=\pi/2$. 
For a Sauter pulse, the dynamically assisted Sauter-Schwinger effect is 
already operational in this regime---which has also been confirmed by 
numerical computations of the total pair-cre\-a\-tion amplitude 
(including the prefactor); see, e.g., Ref.~\cite{Orthaber-Momentum-spectra}.
For a Gaussian profile with the same parameters, on the other hand, 
the weak-field contribution $\propto E_2$ is negligible in the vicinity 
of the instanton trajectory for small values of $\varepsilon$. 
Since the prefactor can be determined by considering small perturbations 
around the instanton trajectory, it is thus also almost unaffected by this 
weak-field contribution $\propto E_2$.
This behavior was confirmed by preliminary numerical investigations 
\cite{Prefactor-footnote} which indicate that the Gaussian prefactor is
approximately constant  (i.e., independent of $\varepsilon$ and $\keld$) for
small $\keld\leq1$ but then starts to grow with both, increasing $\varepsilon$
and $\keld$, for larger values of $\keld$.
By contrast, the prefactor of the Sauter pulse is nearly independent of 
$\varepsilon$ for small as well as large $\keld$ and only varies with 
$\varepsilon$ near threshold $\keld\approx\keld_{\rm crit}=\pi/2$. 
(Note that one cannot simply take the limit $\varepsilon\to0$ here 
since the instanton approximation would break down eventually.) 
In both cases, however, the moderate variation of the prefactor is 
subdominant in comparison to the strong enhancement of the pair-cre\-a\-tion 
probability induced by the changes in the exponent considered here.  

Our results can be generalized in a straightforward manner to other profiles,
for example a Lorentzian profile, which is expected to behave similarly to the
Sauter pulse.
The fact that two visually almost indistinguishable bell-shaped pulses 
(Sauter and Gaussian) display such drastic differences demonstrates that 
tunneling in time-dependent backgrounds can show surprising and partly 
counterintuitive results---which motivates further studies in order to 
understand these nonperturbative phenomena better.


\begin{acknowledgments}
N.S.\ and R.S.\ acknowledge support by Deutsche Forschungsgemeinschaft
(SZ 303/1 and SFB-TR12).
R.S.\ would like to express special thanks to the 
Perimeter Institute for Theoretical Physics 
and the Mainz Institute for Theoretical Physics (MITP)
for hospitality and support.
\end{acknowledgments}

%
%

\appendix

\boldmath

\section{Positions of the singularities $t_{k}^{\star}$}

\unboldmath
The outgoing ratio $R_{k}^{\mathrm{out}}$, which determines
the pair-cre\-a\-tion probability, is approximated by the contour integral
in Eq.~\eqref{eq:contour-integral} according to the WKB method. 
In order to find a suitable
integration contour in the upper complex half-plane $\Im(t)>0$, we
need to know the positions of the integrand's singularities $t_{k}^{\star}$
given by Eq.~\eqref{singularity}. These singularities are poles of the function
$\Xi_{k}(t)$ as well as square-root-type branch points (with associated
cuts) of $\varphi_{k}(t)$. The weak electric field may introduce
further complex singularities into the vector potential $A(t)$ itself and
thus into the integrand; however, this is neither the case for the
Gauss pulse nor for the oscillation.

\subsection{\label{sub:Gauss-pulse-singularities}Gauss pulse}

The dimensionless singularity positions $\tau^{\star}=\omega_{2}t_{k}^{\star}$
are given by Eq.~\eqref{eq:Gauss-pulse-sing-eq} in this case. The regular
solution $\tau_{\mathrm{reg}}^{\star}$
{[}cf.\ Eq.~\eqref{regular-singularity}{]}, which is the only solution to stay
finite in the limit $\varepsilon\to0$, can be Taylor expanded around
$\varepsilon=0$
\begin{multline}
\tau_{\mathrm{reg}}^{\star}=\keld(i-p)-\frac{\sqrt{\pi}}{2}\erf[\keld(i-p)]
\varepsilon+\frac{\sqrt{\pi}}{2}e^{-\keld^{2}(i-p)^{2}}\\
{}\times\erf[\keld(i-p)]\varepsilon^{2}+\ord(\varepsilon^{3})\,\text{.}
\label{eq:gauss-reg-sing-Taylor-series}
\end{multline}
For values of $\keld$ below the threshold
$\keld_{\mathrm{crit}}\approx\sqrt{|\ln\varepsilon|}$,
the terms of order $\ord(\varepsilon)$ and higher are much smaller
than unity. This is no longer true, however, as $\keld$ approaches
$\keld_{\mathrm{crit}}$. Hence, the truncated Taylor
series~\eqref{eq:gauss-reg-sing-Taylor-series}
is appropriate to describe the movement of the regular singularity
as the parameters are varied within the subcritical regime.

\begin{figure}
\subfloat[\label{fig:gauss-sing-varying-gamma}Singularities moving as $\keld$
is varied from $1$ to $7$ (dark red to light yellow dots) for constant
$\varepsilon=10^{-7}$. The associated threshold for the dynamic assistance
is $\sqrt{|\ln\varepsilon|}\approx4$. In the subcritical $\keld$ range,
the regular singularity on the imaginary axis moves essentially as
in the case of the ordinary Sauter-Schwinger effect
($\tau_{\mathrm{reg}}^{\star}\approx i\keld$),
cf.\ Eq.~\protect\eqref{constant}, and the additional singularities are much
deeper in the complex plane. As $\keld$ gets close to the critical threshold,
the regular singularity slows down,
cf.\ Eq.~\protect\eqref{eq:gauss-reg-sing-Taylor-series},
and ``pushes'' the additional singularities away from the real axis.]
{\includegraphics{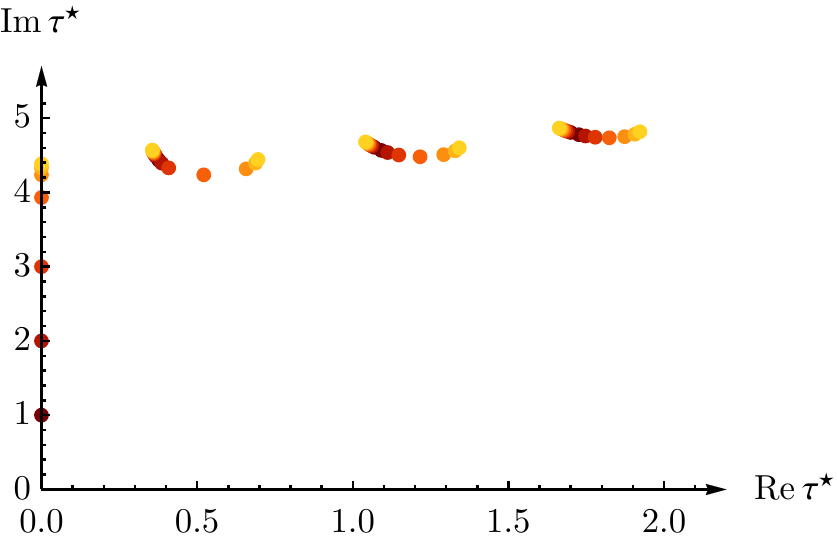}}

\subfloat[\label{fig:gauss-sing-varying-epsilon}Singularities for $\keld=4$
and $\varepsilon=10^{s}$ with $s$ going from $-11$ to $-3$ (dark
red to light yellow dots). The black crosses are the corresponding asymptotic
solutions $\tau_{\mathrm{add},n}^{\star}$ for $n=-1,\ldots,-4$,
see Eq.~\protect\eqref{eq:gauss-add-sing-asymptotic}, which move along the
dashed hyperbolas as $\varepsilon$ is varied. The transition to
the critical regime is at $\varepsilon\approx10^{-7}$. At this value,
the regular singularity on the imaginary axis starts to move towards
the real axis. The additional singularities approach the real axis
continuously as $\varepsilon$ grows. In the subcritical range, the asymptotic
values $\tau_{\mathrm{add},n}^{\star}$ are appropriate to describe
the movement of the additional singularities.]
{\includegraphics{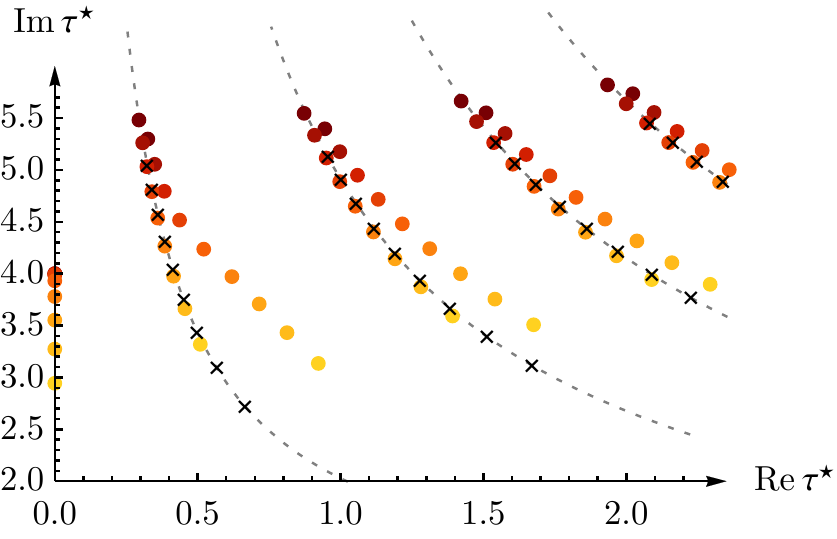}}

\caption{\label{fig:gauss-sing-movement}Numerical solutions
$\tau^{\star}$ of the Gauss-case singularity
equation~\protect\eqref{eq:Gauss-pulse-sing-eq} for the momentum $p=0$.}
\end{figure}

The other solutions of the singularity equation~\eqref{eq:Gauss-pulse-sing-eq}
diverge as $\varepsilon$
approaches zero, so we may approximate the error function by its asymptotic
form $\erf\tau\sim-\exp(\tau^{2})/(\sqrt{\pi}\tau)$~\cite{Abramowitz+Stegun}
for small $\varepsilon$ in order to describe these additional singularities.
In this limit, the $\tau^{\star}$-independent terms on the right-hand
side of Eq.~\eqref{eq:Gauss-pulse-sing-eq} can be neglected, and we find 
the series of singularities
\begin{equation}
\tau_{\mathrm{add},n}^{\star}\approx
\sqrt{\frac{\ln\varepsilon}{\cos(2\theta_{n})}}\;e^{i\theta_{n}}
\label{eq:gauss-add-sing-asymptotic}
\end{equation}
with $n\in\mathbb{Z}$. The complex phase $\theta_{n}$ is the single
real solution of the transcendental equation
\begin{equation}
|\ln\varepsilon|\tan(2\theta_{n})-2\theta_{n}=2\pi n
\label{eq:gauss-add-sing-asymptotic-phase}
\end{equation}
in the range $\pi/4<\theta_{n}<3\pi/4$, which can be found numerically.
Note that, in this approximation, the dependence on momentum $p$ 
and the difference between the plus and the minus cases in 
Eq.~\eqref{eq:Gauss-pulse-sing-eq} disappear.
We thus only get one result where there are actually two different 
singularities ($\pm$ cases). 
For fixed $\varepsilon$, all points $\tau_{\mathrm{add},n}^{\star}$
lie on a hyperbola in the complex plane since
$\Im^{2}(\tau_{\mathrm{add},n}^{\star})-\Re^{2}(\tau_{\mathrm{add},n}^{\star})
=|\ln\varepsilon|$
for all $n$. If we consider a fixed $n$ instead, we can approximate
the solution of Eq.~\eqref{eq:gauss-add-sing-asymptotic-phase} for
sufficiently large $|\ln\varepsilon|$ by linearizing the tangent around
2$\theta_{n}=\pi$. To leading order, this gives the asymptotic form
\begin{equation}
\tau_{\mathrm{add},n}^{\star}\approx-\frac{\pi(n+1/2)}{\sqrt{|\ln\varepsilon|}}
+i\sqrt{|\ln\varepsilon|}\,\text{,}
\end{equation}
which parametrizes another hyperbola as $\varepsilon$ is varied,
cf.\ Fig.~\subref{fig:gauss-sing-varying-epsilon}.

For $p=0$, the singularity equation~\eqref{eq:Gauss-pulse-sing-eq} is symmetric
with respect
to the imaginary axis ($\Re\tau^{\star}\to-\Re\tau^{\star}$) and
the regular singularity remains on this axis for all values of $\varepsilon$
and $\keld$. The movement of the singularities is depicted in
Fig.~\ref{fig:gauss-sing-movement}.

\subsection{\label{sub:App-singularities-osc}Oscillation}

\subsubsection{Graphical solution of the singularity equation}

\begin{figure}
\subfloat[\label{fig:osc-graph-sol-1}Graphical solution of
Eq.~\protect\eqref{eq:osc-sing-eq}.
The function $F(u)$ is plotted for $\keld=8$ (solid, subcritical)
and $\keld=20$ (dashed, supercritical).]
{\includegraphics{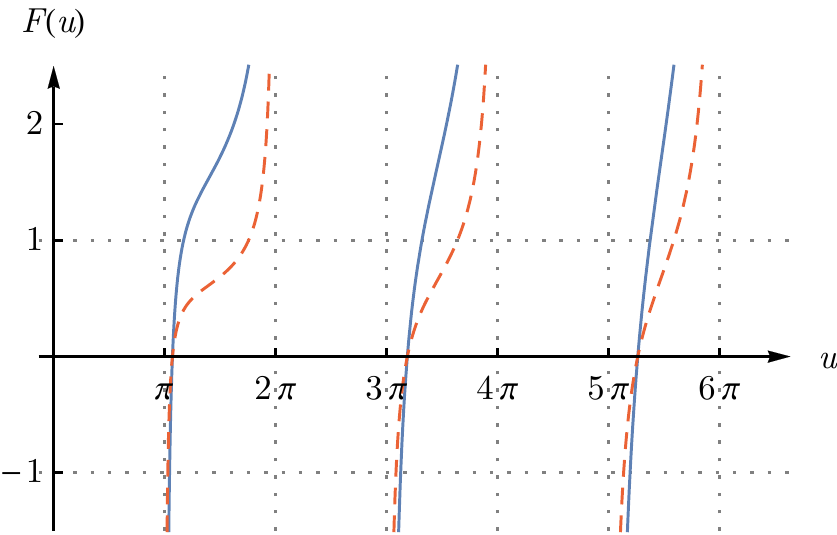}}

\subfloat[\label{fig:osc-graph-sol-2}Imaginary parts of the singularities over
their real parts.]
{\includegraphics{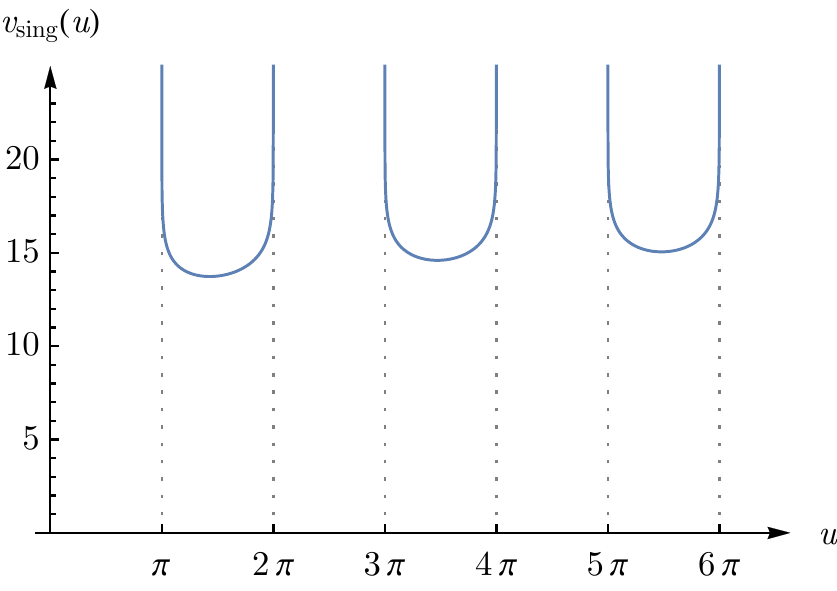}}

\caption{\label{fig:osc-graph-sol}Graphical solution of the singularity
equation~\protect\eqref{eq:oscillation-sing-eq} for $p=0$ and
$\varepsilon=10^{-5}$, so $\keld_\mathrm{crit}$ is roughly
$|\ln\varepsilon|\approx11.5$, cf.\ Eq.~\protect\eqref{critical-Keldysh-cos}.
The function $F(u)$ is even for $p=0$, so we concentrate on positive $u$ for
simplicity.
The real parts $u^{\star}$ of the singularities $\tau^{\star}$ are determined
in the upper plot~(\protect\subref*{fig:osc-graph-sol-1}).
For $p=0$, $F(u>0)$ is only real for $u\in(\pi,2\pi)$, $(3\pi,4\pi)$, etc.\ and
increases strictly over each one of these intervals, respectively, so there are
two singularities per interval.
The corresponding imaginary parts $v^{\star}$ are found by evaluating
$v_{\mathrm{sing}}(u)$ shown in the lower
plot~(\protect\subref*{fig:osc-graph-sol-2}) at $u^{\star}$.}
\end{figure}

The transcendental equation~\eqref{eq:oscillation-sing-eq} for the
dimensionless singularities
$\tau^{\star}$ can be solved graphically. We start to develop the
graphical solution scheme by splitting the equation and the complex
variable $\tau^{\star}=u^{\star}+iv^{\star}$ into real and imaginary
parts
\begin{align}
u^{\star}+\varepsilon\sin(u^{\star})\cosh(v^{\star}) & =
-p\keld\,\text{,}\label{eq:osc-sing-eq-re}\\
v^{\star}+\varepsilon\cos(u^{\star})\sinh(v^{\star}) & =
\pm\keld\,\text{.}\label{eq:osc-sing-eq-im}
\end{align}
For singularities in the upper complex half-plane ($v^{\star}>0$)
with $\sin u^{\star}\neq0$, Eq.~\eqref{eq:osc-sing-eq-re} can be
solved uniquely for the imaginary part $v^{\star}$, which yields
$v^{\star}=v_{\mathrm{sing}}(u^{\star})$ with
\begin{equation}
v_{\mathrm{sing}}(u)=\arcosh\!\left(-\frac{u+p\keld}{\varepsilon\sin u}\right)
\text{.}\label{eq:osc-sing-im-from-re}
\end{equation}
Inserting this expression into Eq.~\eqref{eq:osc-sing-eq-im} leads
to
\begin{equation}
F(u^{\star})=\pm1\label{eq:osc-sing-eq}
\end{equation}
with the function
\begin{multline}
F(u)=\frac{1}{\keld}\left[\arcosh\!\left(-\frac{u+p\keld}{\varepsilon\sin u}
\right)\vphantom{\sqrt{\left(\frac{u+p\keld}{\sin u}\right)^{2}
-\varepsilon^{2}}}\right.\\
\left.\vphantom{\left(-\frac{u+p\keld}{\varepsilon\sin u}\right)}+\cos(u)
\sqrt{\left(\frac{u+p\keld}{\sin u}\right)^{2}-\varepsilon^{2}}\,\right]
\text{.}\label{eq:osc-func-F}
\end{multline}
The real equation~\eqref{eq:osc-sing-eq} can be solved graphically
(see Fig.~\ref{fig:osc-graph-sol}) in order to find all singularities
$\tau^{\star}$ with real parts satisfying $\sin u^{\star}\neq0$.
The corresponding imaginary parts of the singularities are given by
$v_{\mathrm{sing}}(u^{\star})$.

Let us now treat the special case where $\sin u^{\star}=0$, i.e.,
those singularities with real parts $u^{\star}\in\pi\mathbb{Z}$,
which are not covered by the graphical solution scheme. Then,
Eq.~\eqref{eq:osc-sing-eq-re}
reduces to $u^{\star}=-p\keld$, so this case can only occur for
particular momenta $p\in\mathbb{Z}\pi/\keld$. The cosine in
the lower Eq.~\eqref{eq:osc-sing-eq-im} is $\cos u^{\star}=\pm1$
for these singularities. For $\cos u^{\star}=1$ ($u^{\star}=0$,
$\pm2\pi$, etc.), Eq.~\eqref{eq:osc-sing-eq-im} has one unique
solution $v^{\star}>0$ fulfilling $v^{\star}+\varepsilon\sinh v^{\star}=\keld$.
But for $\cos u^{\star}=-1$, in contrast, the left-hand side of
Eq.~\eqref{eq:osc-sing-eq-im}
does not increase strictly with respect to $v^{\star}$, so this equation
can have multiple solutions. It turns out that there are three different
positive solutions $v^{\star}$ for (approximately) subcritical
$\keld<\keld_{\mathrm{crit}}\approx|\ln\varepsilon|$,
see Eq.~\eqref{critical-Keldysh-cos}, and one positive solution for
supercritical $\keld$.
In the case of multiple singularities $\tau^{\star}$ with the same
real value $u^{\star}$, the singularity which lies closest to the
real axis dominates.

\subsubsection{Movement and asymptotics of the singularities}

\begin{figure}
\subfloat[\label{fig:osc-sing-varying-gamma}Singularities for constant
$\varepsilon=10^{-5}$ ($\keld_{\mathrm{crit}}\approx11.5$) and $\keld$ growing
from $7$ to $20$ (dark red to light yellow dots). The linear movement of the
regular singularity $\tau_{\mathrm{reg}}^{\star}\approx i\keld$
(ordinary Sauter-Schwinger effect) slows down exponentially near the
critical threshold according to Eq.~\protect\eqref{eq:osc-zerop-main-sing-eq}.
The additional singularities move along the complex curves
$u+iv_{\mathrm{sing}}(u)$, see Fig.~\protect\subref{fig:osc-graph-sol-2}, which
are independent of $\keld$ for $p=0$.]
{\includegraphics{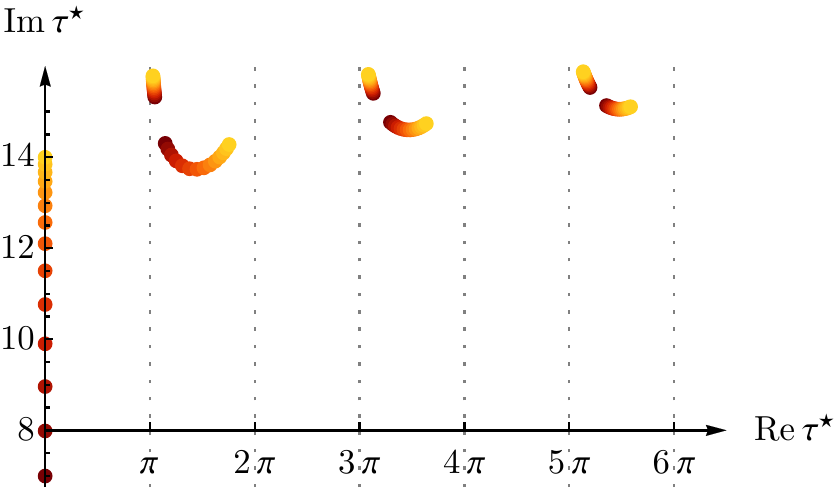}}

\subfloat[\label{fig:osc-sing-varying-epsilon}Moving singularities for
$\keld=13.8$
and $\varepsilon=10^{s}$ with $s=-10,\ldots,-2$ (dark red to light yellow
dots). The critical threshold is at $\varepsilon\approx10^{-6}$. The
regular singularity starts to approach the real axis as the nearby
additional singularities come close. For decreasing $\varepsilon$, each
additional singularity approaches the imaginary axis, converging towards
the edge of its ``$\pi$ interval'',
cf.\ Eq.~\protect\eqref{eq:osc-add-sing-large-epsilon-asymptotic}.]
{\includegraphics{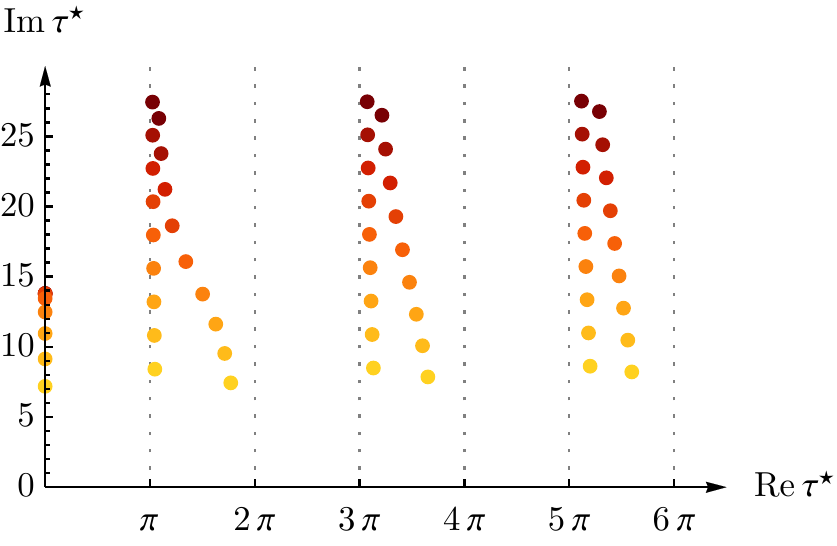}}

\caption{\label{fig:osc-sing-movement}Numerical solutions $\tau^{\star}$
of the singularity equation~\protect\eqref{eq:oscillation-sing-eq} for the
oscillating weak field and $p=0$.}
\end{figure}

In the subcritical regime, the regular singularity is well described
by the Taylor series
\begin{equation}
\tau_{\mathrm{reg}}^{\star}=\keld(i-p)-\sin[\keld(i-p)]\varepsilon+
\sin[2\keld(i-p)]\frac{\varepsilon^{2}}{2}+\ord(\varepsilon^{3})\,\text{.}
\label{eq:osc-reg-sing-Taylor-series}
\end{equation}
As in the Gauss pulse case, the singularity
equation~\eqref{eq:oscillation-sing-eq} can be
solved asymptotically for $p=0$ and $\varepsilon\to0$ (thus
$\Im\tau_{\mathrm{add}}^{\star}\to\infty$)
in order to describe the additional singularities. Here, we get
\begin{equation}
\tau_{\mathrm{add},n}^{\star}\approx\pi(2n-1)+i|\ln\varepsilon|
\label{eq:osc-add-sing-large-epsilon-asymptotic}
\end{equation}
with $n\in\mathbb{Z}$, i.e., each additional singularity converges
against an edge of its ``$\pi$ interval'' for small $\varepsilon$; see
Fig.~\ref{fig:osc-sing-movement}.
Note that each $n$ represents two different solutions of the singularity
equation ($\pm$ cases) again since these two solutions merge asymptotically.

If we want to sum up the contributions from all singularities to the
contour integral~\eqref{eq:contour-integral}, 
we have to approximate the singularities
for fixed parameters $p$, $\varepsilon$, and $\keld$ 
instead of fixed $n$.
Concentrating on the mode $p=0$ for simplicity and inspired by the graphical
method in Fig.~\ref{fig:osc-graph-sol}, we find that the real parts
$u^{\star}$ of the additional singularities do asymptotically
($|u^{\star}|\gg1$) coincide with the centers of the ``$\pi$ intervals'', i.e.,
$\tau_{\mathrm{add}}^{\star}\approx\tau_{\pm,n}^{\star}$ for large
$n\in\mathbb{N}$ with
\begin{align}
\tau_{\pm,n}^{\star} & =\pm\pi\left(2n-\frac{1}{2}\right)+iv_{\mathrm{sing}}\!
\left(\pm2\pi n\mp\frac{\pi}{2}\right)\nonumber \\
 & =\pm\pi\left(2n-\frac{1}{2}\right)+i\arcosh\!
 \left(\frac{2\pi n-\pi/2}{\varepsilon}\right)\text{.}
 \label{eq:osc-add-sing-large-real-parts-asymptotic}
\end{align}
In this context, the $\pm$ sign represents singularities with 
pos\-i\-tive/neg\-a\-tive
real parts. Apart from that distinction, each $n$ corresponds to
two actual singularities again: the $\pm$ cases of Eq.~\eqref{eq:osc-sing-eq},
which merge in the considered limit.
Since $\arcosh x\approx\ln(2x)$ for $x\gg1$, the imaginary part of the 
additional singularities grows logarithmically with $n$ while their 
real part increases linearly.

\section{\label{app:contour-integral}Calculation of the contour
integral~(\protect\ref{eq:contour-integral})}

\begin{figure}
\includegraphics{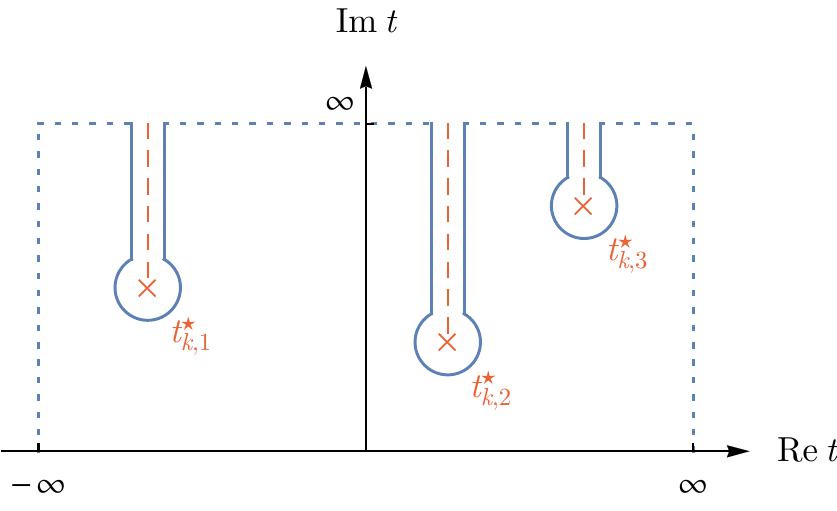}
\caption{\label{fig:int-contour}%
Example for the deformed complex integration contour (dotted parts: no
contribution; solid parts: branch cut contributions and residua) in the case of
three singularities $t_{k,1\text{--}3}^{\star}$ (crosses) with branch cuts
running vertically upwards (dashed lines).%
}
\end{figure}

In order to estimate the integral~\eqref{eq:contour-integral}, the integration
contour can be shifted away from the real axis, upwards into the complex plane,
until the first singularity is met at some $t_{k}^{\star}$. Since
the integrand is analytic for $0<\Im t<\Im t_{k}^{\star}$, this operation
is permitted and the behavior of the complex phase $\varphi_{k}(t)$
gives an estimate of the integrand $\Xi_{k}(t)\exp[2i\varphi_{k}(t)]$
with now complex $t$. However, for the estimation of the integral,
additional control of the integration along the line $\Im t=\Im t_{k}^{\star}$
is required, which is rather difficult to obtain. Therefore, the integration
contour is shifted further upwards, behind the singularities (see
Fig.~\ref{fig:int-contour}), and Cauchy's theorem is applied.

The phase $\varphi_{k}(t)$ and the integrand can be analytically
continued, however, locally on different Riemann sheets (with cuts
in between) as the singularities are of a square-root type. The contribution
from such a shifted integration contour is subdominant due to the
exponential damping coming from the phase
$\varphi_{k}(t)$ {[}Eq.~\eqref{phase-function}{]},
but there remain contributions from integration along both sides of
the cuts and around the singularities. In order to understand how
they can be controlled, it is instructive to consider first the ordinary
Sauter-Schwinger effect as an example.

\subsubsection{Contour integral for a constant electric field $E$}

The advantage of this example is that here contributions from the
singularity and from the cut are explicitly determinable. Since the
singularity $t_{k}^{\star}$ in $\Omega_{k}(t)$---the plus case
of Eq.~\eqref{constant} is relevant in this context---is of square-root type,
the phase $\varphi_{k}(t)$, which is obtained as an integral of $\Omega_{k}(t)$,
behaves as $(t-t_{k}^{\star})^{3/2}$. The residuum (representing
a circle integral around the singularity) can be calculated exactly
\begin{equation}
S_{p}=-\frac{\pi}{2}\exp\left[-\left(\frac{\pi}{2}+i\phi(p)\right)
\frac{E_{S}}{E}\right]
\end{equation}
with the function $\phi(z)$ defined in Eq.~\eqref{eq:Schwinger-phase-aux-func}.
The complex cut emerging from that singularity, chosen vertically upwards from
$t_{k}^{\star}$, produces a jump in the phase function $\varphi_{k}(t)$,
however,
only in its real part. In consequence, the absolute value of the integrand
$\Xi_{k}(t)\exp[2i\varphi_{k}(t)]$ is continuous across the cut and decays
merely polynomially for $\Im t>\Im t_{k}^{\star}$ due to $\Xi_{k}(t)$;
it does not decay exponentially along the cut as might be expected
from the behavior for $\Im t<\Im t_{k}^{\star}$. However, the difference
of values on both sides of the cut oscillates rapidly. The contribution
from the cut can be written as 
\begin{equation}
S_{c}=\exp\left[-\left(\frac{\pi}{2}+i\phi(p)\right)\frac{E_{S}}{E}\right]
\int_{0}^{\infty}I_{c}(u)\,du
\end{equation}
where $I_{c}(u)\approx\sin[2u^{3}E_{S}/(3E)]/u$ is a fast-oscillating
function of $u$. The integral converges solely due to its fast oscillation
and gives approximately the number $\pi/6$. In effect, both contributions
to $R_{k}^{\mathrm{out}}$, 
\begin{equation}
R_{k}^{\mathrm{out}}=S_{p}+S_{c}\approx-\frac{\pi}{3}\exp
\left[-\left(\frac{\pi}{2}+i\phi(p)\right)\frac{E_{S}}{E}\right]\text{,}
\end{equation}
from the singularity and along the cut are of the same order. We take
it as suggestion that neglecting the contribution from the cuts
in more complicated (time-dependent) electric fields should not change
the final result by more than a constant numerical factor.

\subsubsection{Contour integral for the oscillating weak field}

In the case of the os\-cil\-lat\-ing-weak-field profile given by
Eqs.~\eqref{eq:oscillation-E-field}--\eqref{eq:oscillation-potential},
there are infinitely many singularities as shown in
Appendix~\ref{sub:App-singularities-osc}.
Considering $p=0$ again, we get two additional singularities
per ``$\pi$ interval'' (see Fig.~\ref{fig:osc-graph-sol}; the
singularities with $\Re\tau_{\mathrm{add}}^{\star}<0$ are just mirror
images of those with positive real parts) plus the regular singularity
on the imaginary axis. In order to calculate the contour
integral~\eqref{eq:contour-integral},
we need to integrate around all these singularities (generating residua)
and along both sides of the corresponding branch cuts. As the example
above shows, it can be expected that the contributions from the cuts
to $R_{k}^{\mathrm{out}}$ are of the same order as the residua
$S_{p}(t_{k}^{\star})$
at the singularities $t_{k}^{\star}=\tau^{\star}/\omega_{2}$. Hence,
$R_{k}^{\mathrm{out}}$ is approximately given by the residual sum
\begin{equation}
R_{k}^{\mathrm{out}}\approx S_{p}(\tau_{\mathrm{reg}}^{\star})+\sum_{
\tau_{\mathrm{add}}^{\star}}S_{p}(\tau_{\mathrm{add}}^{\star})
\label{eq:complete-residual-sum}
\end{equation}
in dimensionless time units $\tau=\omega_{2}t$. At each singularity,
$\Xi_{k}(t)$ has a pole while $\varphi_{k}(t)$ has merely a square-root-type
branch point, so $\varphi_{k}(t)$ may be treated as constant
when integrating $\Xi_{k}(t)\exp[2i\varphi_{k}(t)]$ along a small
circle around the pole. This way, we get
\begin{equation}
S_{p}(\tau^{\star})=\pm\frac{\pi}{2}e^{2i\varphi_{k}(\tau^{\star})}
\label{eq:residue}
\end{equation}
where the sign depends on whether $\tau^{\star}$ is a plus or minus solution
of the singularity equation~\eqref{eq:oscillation-sing-eq} and on the concrete
potential $A(t)$.
Although we cannot sum the series \eqref{eq:complete-residual-sum} exactly, we
can prove its convergence and estimate its value. Using the asymptotic form of
the singularity positions $\tau_{\mathrm{add}}^{\star}$ for
$|\Re\tau_{\mathrm{add}}^{\star}|\gg1$ and $p=k/m=0$
{[}Eq.~\eqref{eq:osc-add-sing-large-real-parts-asymptotic}{]}, we can
approximate the values of $\varphi_{0}$ at most of the singularities and
estimate the sum over all additional singularities appearing in
Eq.~\eqref{eq:complete-residual-sum}
\begin{equation}
\left|\sum_{\tau_{\mathrm{add}}^{\star}}S_{p}(\tau_{\mathrm{add}}^{\star})
\right|\leq\sum_{\tau_{\mathrm{add}}^{\star}}
|S_{p}(\tau_{\mathrm{add}}^{\star})|=\frac{\pi}{2}
\sum_{\tau_{\mathrm{add}}^{\star}}e^{-2
\Im[\varphi_{0}(\tau_{\mathrm{add}}^{\star})]}\,\text{.}
\end{equation}
Taking into account the contributions from both series of singularities, i.e.,
the $\pm$ cases in Eq.~\eqref{eq:oscillation-sing-eq}, we get
\begin{multline}
\sum_{\tau_{\mathrm{add}}^{\star}}e^{-2\Im[\varphi_{0}(
\tau_{\mathrm{add}}^{\star})]}\approx2\sum_{n=1}^{\infty}\biggl(\exp
\{-2\Im[\varphi_{0}(\tau_{+,n}^{\star})]\}\\
{}+\exp\{-2\Im[\varphi_{0}(\tau_{-,n}^{\star})]\}\biggr)
\end{multline}
with $\tau_{\pm,n}^{\star}$ from
Eq.~\eqref{eq:osc-add-sing-large-real-parts-asymptotic}. For singularities
with large real parts $|\Re\tau^{\star}|\gg1$, we neglect the $m^{2}$ term
under the square root in $\Omega_{k}(t)$ {[}see
Eq.~\eqref{eq:Omega-definition}{]}, and after some other minor analytic
simplifications we get
\begin{equation}
\Im[\varphi_{0}(\tau^{\star})]\approx\frac{E_{S}}{\keld^{2}E_{1}}|
\Re\tau^{\star}|(\Im\tau^{\star}-1)\,\text{.}
\end{equation}
Inserting it into the above sum and performing some further minor
simplifications, we find the estimate
\begin{equation}
\sum_{\tau_{\mathrm{add}}^{\star}}e^{-2
\Im[\varphi_{0}(\tau_{\mathrm{add}}^{\star})]}
<
\frac{4\left(\frac{3\pi E_{1}}{eE_{2}}\right)^{\pi qE_{1}/\omega_{2}^{2}}}
{\left(\frac{3\pi E_{1}}{eE_{2}}\right)^{4\pi qE_{1}/\omega_{2}^{2}}-1}
\label{eq:residual-sum-finite}
\end{equation}
in terms of the usual physical quantities. The right-hand side is
finite, which proves the convergence of the residual sum (at least
for $p=0$).
As a check, two limiting cases can be considered:
\begin{itemize}
\item[a)] $E_{2}\to0$: 
As the assisting oscillation vanishes, we expect the 
ordinary Sauter-Schwinger effect.
Consistently, the right-hand side of Eq.~\eqref{eq:residual-sum-finite}
becomes zero in this limit---there is no contribution from the
additional singularities.
\item[b)] $\omega_{2}^{2}\to0$: 
If the oscillation becomes a static field which has a negligible 
influence due to $E_{2}\ll E_{1}$, the right-hand side of 
Eq.~\eqref{eq:residual-sum-finite} vanishes again, 
leaving the ordinary Sauter-Schwinger effect, as expected.
\end{itemize}

%
%



\begin{thebibliography}{99}
\bibitem{Bethe-Heitler}
There are also studies of scenarios where pair creation is assisted by
nuclear fields (see, e.g., Ref.~\cite{Bethe-Heitler-examples}) but we do not
consider this case here.
\bibitem{Zahn-footnote}
Note that one should carefully distinguish between the perturbative effect of
vacuum polarization during the pulse---which can be interpreted as a suitably
defined intermediate particle number---and the nonperturbative
Sauter-Schwinger effect corresponding to the number of created particles after
the pulse; see also Ref.~\cite{Zahn}.
\end{thebibliography}
\end{document}